\documentclass[a4paper,twocolumn,english,prl,superscriptaddress]{revtex4}
\usepackage[T1]{fontenc}
\usepackage[latin1]{inputenc}
\usepackage{psfrag}

\makeatletter

\usepackage{graphicx} \usepackage{amsmath} \usepackage{amssymb}
\usepackage{babel} \makeatother

\begin{document}

\title{Spontaneous parity breaking of graphene in the quantum Hall regime}

\author{Jean-No\"el~Fuchs}

\author{Pascal~Lederer}
\affiliation{Laboratoire de Physique des Solides, Univ. Paris-Sud,
CNRS, UMR 8502, F-91405 Orsay, France.}

\begin{abstract}
We propose that the inversion symmetry of the graphene honeycomb
lattice is spontaneously broken via a magnetic field dependent
Peierls distortion. This leads to valley splitting of the $n=0$
Landau level but not of the other Landau levels. Compared to quantum
Hall valley ferromagnetism recently discussed in the literature,
lattice distortion provides an alternative explanation to all the
currently observed quantum Hall plateaus in graphene.
\end{abstract}

\date{\today}
\maketitle

Recent experiments have revealed peculiar quantum Hall (QH) effects
in graphene, a single atomic layer of graphite
\cite{Novoselov,Zhang1}. These measurements are understood as single
electron effects and unusual QH features can be traced back to the
relativistic-like dispersion relation of electrons in graphene and
to their twofold valley degeneracy. In particular, the observed
plateaus in the Hall conductivity at filling factor
$\nu=\pm2;\pm6;\pm10$ can be easily understood in this framework
\cite{SchakelZhengGusyninPeres}. Following these pioneering
experiments, Zhang \emph{et. al.} \cite{Zhang2} discovered new QH
plateaus at $\nu=0;\pm1;\pm4$, which several authors
\cite{Nomura,Abanin,Fertig,Goerbig,Alicea} attribute to valley (and
spin) ferromagnetism, relying on interactions between electrons.
However, the absence of plateaus at $\nu=\pm3;\pm5$ is intriguing in
this respect and cast doubts on this interpretation. Alicea and
Fisher \cite{Alicea} propose that disorder might be so strong in
current graphene samples as to destroy exchange interactions and
therefore ferromagnetism~\cite{Nomura}. If this is indeed the case,
one still has to explain the origin of the extra plateaus. Alicea
and Fisher suggest a valley splitting mechanism relying essentially
on lattice scale electron repulsion while neglecting exchange
interactions in a dirty graphene sample \cite{Alicea}. Another
scenario relying on long-range electron interactions and leading to
an excitonic valley gap is the so-called ``magnetic catalysis''
\cite{Magneticcatalysis}. In the present paper, we take a different
route and assume from the outset that direct interactions between
electrons do not play a major role. We explore the possibility that
all the plateaus observed so far could be understood as integer QH
states resulting from electron-lattice interaction effects. The new
input of our model is a magnetic field driven out-of-plane lattice
distortion lifting the valley degeneracy.

Graphene is a honeycomb lattice of carbon atoms: a two dimensional
triangular Bravais lattice with a basis of two atoms, usually
referred to as A and B. The distance between nearest neighbor atoms
is $a=0.14$~nm and the lattice constant is $a\sqrt{3}$.
Experimentally, graphene sheets of area $\mathcal{A} \sim (3 - 10\,
\mu$m$)^2$ are deposited on SiO$_2$/Si substrate. Applying a gate
potential $V_g$ via the substrate allows one to control the
electronic filling of the graphene bands. The number of induced
electronic charges is given by $N_c=V_g C_g/e$ where the capacitance
per unit area can be estimated as $C_g/\mathcal{A}\approx \epsilon_r
\epsilon_0 /d \approx 1.2\times 10^{-4}$~F/m$^2$, where $-e<0$ is
the electron charge, $\epsilon_r \approx 4$ is the silicon oxide
dielectric constant and the thickness $d\sim 300$~nm
\cite{Novoselov,Zhang1}.

In order to study the electronic properties of graphene, we use a
standard nearest neighbor tight-binding model \cite{Wallace} with
hopping amplitude $t\approx 3$~eV \cite{Saito}. It describes the
hopping of electrons between $2p_z$ carbon orbitals. There is one
electron per carbon atom. If we call $N_p$ the number of plaquettes
(or unit cells), there are $2N_p$ electrons in the sample under zero
gate potential. The first Brillouin zone is hexagonal and of its six
corners, only two are inequivalent and usually called $K$ and $K'$.
We choose $\mathbf{K}=4\pi/(3\sqrt{3}a)\mathbf{u}_x$ and
$\mathbf{K'}=-\mathbf{K}$. The resulting band structure features the
merging of the conduction and valence band at precisely these two
points: graphene is a two valley ($K$ and $K'$) zero-gap
semiconductor. Near these so-called Dirac points, the electrons
behave as charged massless Weyl (or chiral Dirac) fermions with
Fermi velocity $v_F=3at/2\hbar \approx 10^6$~m/s playing the role of
an effective light velocity in the relativistic-like dispersion
relation $\varepsilon_k=\pm \hbar v_F |\mathbf{k}|$. When the gate
voltage is zero, the ``big band'' (valence plus conduction band) is
half-filled: the Fermi level is right at the Dirac points.

Adding a weak perpendicular magnetic field $B_\perp$, such that the
flux per plaquette is much smaller than the flux quantum
$\phi_0=h/e$, McClure first obtained the relativistic Landau levels
(LL) of graphene \cite{McClure}, see also
Ref.~\cite{Semenoff,Haldane}. Including the Zeeman effect, the LL in
the Dirac equation approximation read
\begin{equation}
\varepsilon_{n,\sigma}=\text{sgn}(n)\sqrt{|n|}\hbar \omega_c +
\frac{g^*}{2}\mu_B B_\text{tot} \sigma \, , \label{LL1}
\end{equation}
where the ``cyclotron energy'' is $\hbar \omega_c=\sqrt{2}\hbar
v_F/l_B$, the LL index $n$ is an integer, the spin projection along
the magnetic field axis is $\sigma=\pm 1$, the Bohr magneton is
$\mu_B=e\hbar/2m$, with $m$ the bare electron mass, and the
effective $g$-factor is $g^*\approx 2$ close to its bare value
\cite{Zhang2}. The magnetic length is defined as usual by
$l_B=\sqrt{\hbar/eB_\perp}$. If the Zeeman splitting is negligible,
each LL has degeneracy $4N_\phi$. The total number of flux quanta
across the sample $N_\phi=B_\perp \mathcal{A}/\phi_0$ gives the
orbital degeneracy. The factor $4$ accounts for spin $1/2$ and
twofold valley degeneracy. We call $\nu=N_c/N_\phi=C_g V_g/e N_\phi$
the filling factor. When the gate voltage is zero, $\nu=0$ and the
$n=0$ (central) Landau level (CLL) is half-filled as a result of
particle-hole symmetry leading to the remarkable fact that the
number of electrons in the CLL is $2N_\phi$.

We now consider a spontaneous out-of-plane lattice distortion that
-- in presence of a substrate -- breaks the inversion symmetry of
the honeycomb lattice and provides a mechanism for lifting the
valley degeneracy. Assume that the A (resp. B) sublattice moves away
(resp. towards) the substrate by a distance $\eta$ \footnote{Here we
assume that the average distance of the graphene sheet to the
substrate is fixed. Relaxing this constraint would lower the total
energy even further by providing an extra variational parameter.}.
Electrons are still described by a honeycomb nearest neighbor
tight-binding model, however the two atoms in the basis now have
different on-site energies \footnote{The hopping amplitude is also
slightly modified by the distortion but is still unique. The only
significant modification is the appearance of two different on-site
energies. Here we neglect the distortion induced change in hopping
amplitude, as the distortion is very small.}. The energy on atom A/B
is called $\pm M$ following Haldane \cite{Haldane}, who calculated
the LL of such a system. Close to the Dirac points, it reads
\begin{eqnarray}
\varepsilon_{n,\sigma,\alpha}&=&\text{sgn}(n)\sqrt{M^2+2\hbar v_F^2
eB_\perp |n|} \nonumber \\
&+& \frac{g^*}{2}\mu_B B_\text{tot} \sigma  \text{ if } n \neq 0 \label{LL2}\\
\varepsilon_{0,\sigma,\alpha}&=&\alpha M+ \frac{g^*}{2}\mu_B
B_\text{tot} \sigma \text{ if } n=0\, , \label{LL3}
\end{eqnarray}
where $\alpha = \pm 1$ is the valley index corresponding to the
Dirac points $\alpha \mathbf{K}$. In terms of the low-energy
effective theory, the distortion means that the Weyl fermions
spontaneously acquire a finite mass. Note that the on-site energy
difference lifts valley degeneracy for the CLL \emph{only}. In
addition the effect of a nonzero on-site energy $M$ on each $n\neq0$
LL is very weak, of order $M^2/\hbar v_F^2eB_\perp \sim 5.10^{-4}$
for a typical magnetic field $\sim 10$~T as we will see. We could
therefore set $M=0$ in the $n\neq 0$ LL and use the approximate
Eq.~(\ref{LL1}) instead of Eq.~(\ref{LL2}). However, we shall see
below that in order to compute the lattice distortion it is
important to keep Eq.~(\ref{LL2}).

Such a lattice distortion spontaneously occurs because it lowers the
total energy, in a way similar to Peierls's mechanism \cite{Peierls}
except for the magnetic field playing an essential role here and for
the crystal being two rather than one dimensional. Assume that the
last partially filled LL is $n=0$ (i.e. the gate voltage $V_g$ is
such that $|\nu| \leq 2$). We show that in this case it is always
favorable to slightly distort the lattice \emph{provided} there is a
perpendicular magnetic field \footnote{When $|\nu|>2$, there is no
energy gained by distortion. The distortion only occurs for $\nu$
close to zero.}. The distortion lowers the electronic energy. This
energy lowering comes both from the CLL, which gives an essential
contribution, and also from all the $n<0$ LL, which contribute in a
less important way as we explain below. There are $(2+\nu)N_\phi$
electrons in the CLL. They contribute an energy gain
\begin{equation}
E_{n=0}=-N_\phi (2-|\nu|)M \label{electron}
\end{equation}
because when $\nu<0$, all $(2+\nu)N_\phi$ electrons gain each an
energy $M$ and when $\nu>0$, $2N_\phi$ electrons gain each an energy
$M$ but the remaining $\nu N_\phi$ electrons loose each an energy
$M$. This energy gain depends on the magnetic field through
$N_\phi$. In addition, the energy gain is linear in the out-of-plane
distortion $\eta$ because the on-site energy is proportional to the
distortion, as we discuss below: $M = D \eta$, where $D$ is a
proportionality constant, akin to a deformation potential. The other
$2(N_p-N_\phi)$ electrons that fill the $n<0$ LLs, also contribute
to the energy lowering. Each of them gains a small energy compared
to what an $n=0$ electron gains, as discussed in the preceding
paragraph, but as there are many more of them, about $2(N_p-N_\phi)
\approx 2N_p$ , we can not neglect their contribution. In the Dirac
equation approximation, we find
\begin{equation}
E_{n<0}=-\gamma \frac{N_p a}{\hbar v_F}M^2 \, , \label{electron2}
\end{equation}
where the numerical factor $\gamma =3^{1/4}/\sqrt{\pi} \approx 0.74$
\footnote{The Dirac equation approximation is strictly valid only
for LL such that $|n|\ll N_p/2N_\phi$. In the full tight-binding
model, $E_{n<0}$ has the same structure albeit with a slightly
smaller numerical factor $\gamma \approx 0.67$, see p. 1810 in
\cite{Saito}.}.
This energy gain is quadratic in the distortion, and therefore
smaller than $E_{n=0}$ at small distortion, and independent of the
magnetic field. Actually, this term represents the full electronic
energy gain for a lattice distortion under \emph{zero} magnetic
field. In the end, adding $E_{n<0}$ to $E_{n=0}$, we see that the
larger the magnetic field, the larger the electronic energy gain.

The distortion costs an elastic energy
\begin{equation}
E_{\text{elastic}}=N_p G \eta^2 \label{elastic}\, ,
\end{equation}
where the out-of-plane distortion is assumed to be small $\eta \ll
a$ and $G$ is an elastic constant. As $E_{n<0}$ and
$E_{\text{elastic}}$ are both quadratic in the lattice distortion,
we introduce a renormalized elastic constant $G'=G-\gamma aD^2/\hbar
v_F$ and write an effective elastic energy:
\begin{equation}
E_{\text{elastic}}+E_{n<0}=N_p G' \eta^2 \label{elastic2}\, .
\end{equation}
The effect of the $n<0$ electrons is to reduce the lattice stiffness
and therefore to enhance the distortion. We take it as an
experimental fact that there is no spontaneous out-of-plane
distortion in absence of perpendicular magnetic field, see also
\cite{Saito}, which means that $G'>0$ \footnote{Indeed, the total
shift of the valence band energy when $B=0$ is identical to
$E_{n<0}$ when $B\neq 0$.}.

We now estimate the two constants $D$ and $G$. From the frequency
$\omega_0/2\pi c \sim 800$~cm$^{-1}$ of the \emph{graphite}
out-of-plane optical phonon \cite{Dresselhaus}, we obtain $Ga^2
\approx m_c \omega_0^2a^2/4 \sim 14$~eV, where $m_c$ is the carbon
atom mass \footnote{Measuring the corresponding phonon mode in
\emph{graphene on substrate} would directly give access to $G'$, and
therefore provide an independent determination of the constant $D$
through the equation $D=\sqrt{(G-G')\hbar v_F/\gamma a}$.}. The
condition $G'>0$ then implies that $Da<\sqrt{Ga\hbar v_F
/\gamma}\approx 9.8$~eV. The experiment \cite{Zhang2} suggests that
valley splitting is larger than Zeeman splitting, which occurs in
our model if $Da \gtrsim 6.3$~eV, as we show below. It is quite
difficult to accurately predict the constant $D$ and we will
therefore only provide an order of magnitude estimate. The mechanism
that we think gives the largest contribution results from the
interaction of a single carbon atom with the SiO$_2$ substrate
treated as a dielectric continuum \footnote{For simplicity, we do
not take the atomic structure of the SiO$_2$ layer into account and
leave it for further studies.}. The non-retarded Lennard-Jones
interaction energy of an atom at a distance $r$ of a dielectric wall
is given by $E_\text{LJ}(r)\approx -(\epsilon_r-1)\langle
\mathbf{d}^2\rangle/(\epsilon_r+1)48\pi \epsilon_0 r^3$, where
$\langle \mathbf{d}^2\rangle$ is the atomic ground state expectation
value of the squared electric dipole moment \cite{Aspect}. The
on-site energy change resulting from the lattice distortion may be
estimated as
\begin{equation}
\pm M \approx E_\text{LJ}(d_0\pm \eta)-E_\text{LJ}(d_0)\approx \pm
\frac{\epsilon_r-1}{\epsilon_r+1}\frac{\langle
\mathbf{d}^2\rangle}{16\pi \epsilon_0 d_0^4} \eta
\end{equation}
where the $\pm$ sign refers to sublattice A ($+1$) or B ($-1$)
\footnote{One should not mistake the sublattice index $l=\pm 1$ (A
or B) for the valley index $\alpha=\pm 1$ ($K$ or $K'$). These
indices are only equivalent in the CLL.}, $d_0$ is the average
distance between the graphene sheet and the substrate and we assumed
that $\eta \ll d_0$. For a carbon atom $\sqrt{\langle
\mathbf{d}^2\rangle} \sim 4ea_0$, where $a_0$ is the Bohr radius,
which gives $Da\sim
a(\epsilon_r-1)e^2a_0^2/(\epsilon_r+1)\pi\epsilon_0 d_0^4\sim 1$ to
$14$~eV depending on $d_0\sim$ $0.1$ to $0.2$~nm. Therefore, the
order of magnitude of the deformation potential $Da$ is $5$~eV. From
now on, in order to match the experiment \cite{Zhang2}, we take the
plausible value $Da=7.8$~eV, which gives $G'a^2\approx 4.2$~eV.

Minimizing
$E_\text{tot}=E_{\text{n=0}}+E_{\text{n<0}}+E_{\text{elastic}}$ as a
function of the distortion $\eta$, we obtain an on-site energy
\begin{equation}
M=D\eta =\frac{N_\phi}{N_p} \frac{2-|\nu|}{2} \frac{D^2}{G'}\, ,
\label{Peierls}
\end{equation}
and a condensation energy $E_\text{tot}=-(2-|\nu|)N_\phi M/2$. The
distortion is indeed very small, of order $\eta/a \sim
2.10^{-5}\times B_\perp [\text{T}]$ when $\nu \approx 0$. This gives
an $n=0$ valley splitting $\Delta_v = 2M \approx 4.2\text{K} \times
(1-|\nu|/2)B_\perp [\text{T}]$, which for $\nu \approx 0$ is larger
than the Zeeman splitting $\Delta_Z =g^* \mu_B B_\text{tot} \approx
1.5\text{K} \times B_\text{tot}[\text{T}]$ \footnote{The valley
splitting is very sensitive to the precise value of $D$ because it
is proportional to $D^2/G'$ and diverges as $Da$ reaches
$\sqrt{Ga\hbar v_F /\gamma}$.}. The on site energy $M$ is indeed
much smaller than the cyclotron energy and can therefore be safely
neglected in each $n\neq 0$ LL: $M/\hbar \omega_c \sim
5.10^{-3}\times \sqrt{B_\perp[T]}$ when $\nu \approx 0$. This means
that the LL spectrum for $n\neq 0$ is approximately given by
Eq.~(\ref{LL1}) -- as in the case of no lattice distortion -- and
therefore $\varepsilon_{n,\alpha} \propto \sqrt{B_\perp}$ in
agreement with recent spectroscopic observations \cite{Sadowski}.

Considering LL broadening due to disorder, the preceding calculation
for lattice distortion is modified at weak magnetic field, when the
valley splitting $\Delta_v$ is smaller than the LL width
$\Delta_\text{imp}$. For example, for rectangular LL -- the density
of states being $4N_\phi/\Delta_\text{imp}$ inside a LL and zero
otherwise -- Eq.~(\ref{electron}) is changed into
\begin{equation}
E_{n=0}=-\frac{2N_\phi}{\Delta_\text{imp}}M^2 =
-\frac{2N_\phi}{\Delta_\text{imp}}D^2 \eta^2 \, ,
\end{equation}
while Eq.~(\ref{electron2}) remains unchanged since it concerns
totally filled LLs. The electronic energy gain is now proportional
to $\eta^2$. Comparing this energy to the renormalized elastic
energy loss of Eq.~(\ref{elastic2}), we see that a distortion only
occurs if $2N_\phi D^2/\Delta_\text{imp}>N_p G'$, which always
happen at large enough magnetic field. This condition is precisely
equivalent to requiring that the valley splitting $\Delta_v=2M$ --
given by Eq.~(\ref{Peierls}) with $\nu \approx 0$ -- be larger than
the LL width $\Delta_\text{imp}$. This is satisfied if
$B_\perp>hG'\Delta_\text{imp}/3\sqrt{3}ea^2D^2\sim 7$~T, where we
used $\Delta_\text{imp}=2\Gamma$ with $\Gamma \sim 15$~K the
measured LL half-width at half-maximum \cite{Zhang2}. Therefore, as
soon as $B_\perp$ is larger than this threshold value, the lattice
is distorted and the valley gap is larger than the LL width, which
means that one can use the results obtained in the preceding
paragraph in the case of infinitely narrow LL.

\begin{figure}[ptb]
\psfrag{4}{$4$} \psfrag{2}{$2$} \psfrag{1}{$1$} \psfrag{B}{$B$}
\psfrag{0}{$0$} \psfrag{epsilon}{$\varepsilon$} \psfrag{n=1}{$n=1$}
\psfrag{n=-1}{$n=-1$} \psfrag{n=0}{$n=0$} \psfrag{omegac}{$\hbar
\omega_c$} \psfrag{deltav}{$\Delta_v$} \psfrag{deltaz}{$\Delta_Z$}
\psfrag{a}{$(1,1,\alpha)$} \psfrag{b}{$(1,-1,\alpha)$}
\psfrag{c}{$(0,1,1)$} \psfrag{d}{$(0,-1,1)$} \psfrag{e}{$(0,1,-1)$}
\psfrag{f}{$(0,-1,-1)$} \psfrag{g}{$(-1,1,\alpha)$}
\psfrag{h}{$(-1,-1,\alpha)$}
\includegraphics[width=5.5cm]{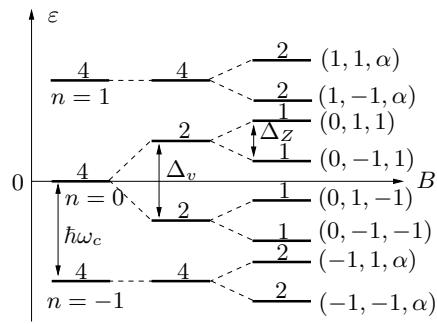}
\caption{Energy $\varepsilon$ of the first LL versus magnetic field
$B$. The degeneracy in units of the flux number $N_\phi$ appears on
the levels. The cyclotron $\hbar \omega_c$, valley $\Delta_v$ and
Zeeman $\Delta_Z$  gaps are also specified. At large $B$, the levels
are tagged by the LL $n$, spin $\sigma$ and valley $\alpha$ indices:
$(n,\sigma,\alpha)$.} \label{figureLL}
\end{figure}
We now discuss the expected plateaus in the Hall conductivity
$\sigma_{xy}=\nu e^2/h$ as a function of the filling factor $\nu
\propto V_g$ and the magnetic field. We consider a system at low
temperature $T\ll \Delta_\text{imp}$, where thermal effects can be
neglected, and assume broadened LL with a width
$\Delta_\text{imp}\sim 30$~K that we compare to the calculated gaps:
for typical magnetic fields, the cyclotron gap $\hbar
\omega_c\approx 420\text{K} \times \sqrt{B_\perp[\text{T}]}$ is the
largest, then the valley gap is $\Delta_v\approx 4.2\text{K} \times
(1-|\nu|/2)B_\perp[\text{T}]$ and finally the Zeeman gap
$\Delta_Z\approx 1.5\text{K} \times B_\text{tot}[\text{T}]$ is the
smallest, see Figure \ref{figureLL}. When the magnetic field is such
that the cyclotron gap becomes larger than $\Delta_\text{imp}$ which
occurs at $B_\perp \sim 5.10^{-3}$~T, one expects plateaus at
$\nu=\pm (4|n|+2)$. Then, when the valley gap (corrected by Zeeman
splitting) $\Delta_v-\Delta_Z$ becomes of order $\Delta_\text{imp}$
which occurs at $B_\perp \sim 11$~T for $\nu\approx 0$ -- thanks to
our choice of $Da$ -- one expects a $\nu=0$ plateau. Finally, when
the Zeeman gap reaches $\Delta_\text{imp}$ which occurs at
$B_\text{tot}\sim 20$~T, one expects plateaus at $\nu=\pm 1$ and
$\nu=4n$ ($n\neq 0$). Because valley degeneracy is not lifted by the
lattice distortion when $n\neq 0$, plateaus are not expected at
$\nu=\pm (2|n|+1)$ ($n\neq 0$). Experimentally, plateaus at $\nu=\pm
2; \pm 6; \pm 10$ are observed at magnetic field $\sim 9$~T
\cite{Novoselov,Zhang1} and are attributed to the cyclotron gap, the
$\nu=0$ plateau appears at $11$~T, $\nu=\pm 1$ and $\pm 4$ are
observed at $B_\perp >17$~T and the $\nu=\pm 3; \pm 5$ plateaus are
not observed \cite{Zhang2}. This agrees qualitatively with our model
and allows one to attribute the $\nu=0$ plateau to the $n=0$ valley
gap and the $\nu=\pm1;\pm4$ plateaus to the Zeeman gap.

The $\nu=0$ plateau, which occurs around zero gate voltage, is worth
considering from an edge states perspective \cite{Halperin}. We
assume smooth edges on the sides of a sample of width $W$ and take
infinite mass confinement as boundary condition, following
Ref.~\cite{Tworzydlo}. The on-site energy (the ``mass'') is now
position dependent in the $y$ direction perpendicular to the edges:
in the bulk, $M(y)$ is constant and given by Eq.~(\ref{Peierls}); on
the edges $y\approx \pm W/2$, it smoothly rises to infinity in order
to confine the electrons. Eq.~(\ref{LL2}) and (\ref{LL3}) show that
electronic states with positive (resp. negative) energy bend upward
(resp. downward) in energy on the edges as $M(y)\to \infty$. As
$\Delta_v>\Delta_z$, the sign of the energy is given by that of the
LL index $n$ except for $n=0$ where it is given by the valley index
$\alpha$. Therefore when the Fermi level lie in the valley gap,
there are \emph{no} edge states, and the Hall conductivity
$\sigma_{xy}=0$, as expected. The absence of edge states (when $\nu
\approx 0$) is a consequence of the valley splitting being larger
than the Zeeman splitting, see Ref.~\cite{Abanin,Fertig}. At the
same time the longitudinal conductivity $\sigma_{xx}$ should be
exponentially small (activated) because of the absence of current
carrying states \emph{both} in the bulk and on the edges: therefore,
one does not expect a wide zero in the longitudinal resistivity
$1/\sigma_{xx}$, as for usual QH states, but rather in the
conductivity $\sigma_{xx}$. Actually, the system should conduct as a
very bad metal, which according to Mott's criterion implies
$\sigma_{xx}\sim e^2/h$, just as for graphene under zero magnetic
field \cite{Novoselov,Zhang1}. This point deserves further studies.
In the experiment \cite{Zhang2}, when a $\nu=0$ plateau is observed
in $\sigma_{xy}$ at $25$~T, the longitudinal resistance features a
finite peak $R_{xx}\sim 40$~k$\Omega$, corresponding to a
resistivity of order $1/\sigma_{xx}\sim 10$~k$\Omega$ of the same
order as that measured at zero magnetic field $1/\sigma_{xx}\approx
h/4e^2\approx 6.5$~k$\Omega$ \cite{Novoselov,Zhang1}.

In conclusion, we compare the predictions of our model to that of
valley ferromagnetism \cite{Nomura,Abanin,Fertig,Goerbig,Alicea}.
First, we predict that valley degeneracy is not lifted in $n\neq 0$
LL, whereas valley ferromagnetism lifts this degeneracy. This
results in the absence of the $\nu=\pm (2|n|+1)$ plateaus, with
$n\neq 0$. Second, the valley gap is proportional to the
perpendicular magnetic field, whereas the $n=0$ skyrmion gap
relevant for ferromagnetism $\Delta_\text{sky}\sim e^2/\epsilon l_B$
scales as $\sqrt{B_\perp}$ \cite{Goerbig,Alicea}: this should be
seen in activation gaps measurements \cite{Zhang2}. In addition,
using the coincidence method with a tilted magnetic field
\cite{Zhang2}, one should be able to distinguish the different gaps
through their dependence in the perpendicular or total magnetic
field. The gate voltage dependence of the valley gap
$\Delta_v\propto (1-|\nu|/2)$ could be detected spectroscopically
\cite{Sadowski}. Third, if lattice distortion indeed occurs it
should be directly seen. It might be detected using synchrotron
X-ray diffraction at grazing incidence or scanning tunneling
microscopy at magnetic fields $\sim 10$~T and low temperature.
Fourth, the lattice distortion and its consequences should vanish if
the graphene sheet is placed in a symmetric dielectric environment.
In the end, we provide what we think is a plausible mechanism for
lifting valley degeneracy. Whether lattice distortion indeed occurs
remains to be checked experimentally.

We thank M.~Goerbig, R.~Moessner, F.~Pi\'echon, Ch.~Texier and the
other participants in the ``graphene journal club'' in Orsay for
many useful discussions.

\end{document}